\documentclass[fleqn,usenatbib]{mnras}

\usepackage{newtxtext,newtxmath}

\usepackage[T1]{fontenc}
\usepackage{ae,aecompl}
\usepackage{soul}

\usepackage{graphicx}	
\usepackage{amsmath}	
\usepackage{amssymb}	
\usepackage[usenames]{xcolor}
\usepackage{natbib}
\usepackage{hyperref}
\usepackage{xfrac}

\def\equationautorefname~#1\null{equation~(#1)}

\DeclareMathAlphabet{\mathpzc}{OT1}{pzc}{m}{it}\definecolor{purple}{RGB}{160,32,240}

\newcommand{\Msun}{\,{\rm M}_{\odot}}

\newcommand{\pc}{\,{\rm pc}}

\newcommand{\Mstar}{M_{\star}}

\newcommand{\Mc}{M_{\rm c}}

\newcommand{\sigcl}{\sigma_{\rm cloud}}
\newcommand{\sigb}{\sigma_{\rm clump}}
\newcommand{\Mcl}{M_{\rm cloud}}
\newcommand{\Rcl}{R_{\rm cloud}}
\newcommand{\Rb}{R_{\rm clump}}
\newcommand{\Mb}{M_{\rm clump}}
\newcommand{\avir}{\alpha_{\rm vir}}
\newcommand{\Rstar}{R_{\star}}
\newcommand{\Rgc}{R_{\rm GC}}

\newcommand{\sigg}{\Sigma_{\rm g}}
\newcommand{\phiPbar}{\phi_{\bar{P}}}

\title[The stellar cluster mass-radius relation]{On the initial mass-radius relation of stellar clusters}

\author[Choksi \& Kruijssen]{Nick Choksi$^{1,2 \href{https://orcid.org/0000-0003-0690-1056}{\includegraphics[scale=0.4]{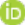}}}$\thanks{E-mail: nchoksi@berkeley.edu} and
J.~M.~Diederik Kruijssen$^{2\href{https://orcid.org/0000-0002-8804-0212}{\includegraphics[scale=0.4]{sizes/orcid.pdf}}}$
\\
$^{1}$Department of Astronomy and Theoretical Astrophysics Center, University of California Berkeley, Berkeley, CA 94720\\
$^{2}$Astronomisches Rechen-Institut, Zentrum f\"{u}r Astronomie der Universit\"{a}t Heidelberg, M\"{o}nchhofstra\ss e 12-14, 69120 Heidelberg, Germany
}

\date{Accepted XXX. Received YYY; in original form 2021 July 4}

\pubyear{2021}

\begin{document}
\label{firstpage}
\pagerange{\pageref{firstpage}--\pageref{lastpage}}
\maketitle

\begin{abstract}
Young stellar clusters across nearly five orders of magnitude in mass appear to follow a power-law mass-radius relationship (MRR), $\Rstar \propto \Mstar^{\alpha}$, with $\alpha \approx 0.2 - 0.33$. We develop a simple analytic model for the cluster mass-radius relation. We consider a galaxy disc in hydrostatic equilibrium, which hosts a population of molecular clouds that fragment into clumps undergoing cluster formation and feedback-driven expansion. The model predicts a mass-radius relation of $\Rstar \propto \Mstar^{1/2}$ and a dependence on the kpc-scale gas surface density $\Rstar \propto \sigg^{-1/2}$, which results from the formation of more compact clouds (and cluster-forming clumps within) at higher gas surface densities. This environmental dependence implies that the high-pressure environments in which the most massive clusters can form also induce the formation of clusters with the smallest radii, thereby shallowing the observed MRR at high-masses towards the observed $\Rstar \propto \Mstar^{1/3}$. At low cluster masses, relaxation-driven expansion induces a similar shallowing of the MRR. We combine our predicted MRR with a simple population synthesis model and apply it to a variety of star-forming environments, finding good agreement. Our model predicts that the high-pressure formation environments of globular clusters at high redshift naturally led to the formation of clusters that are considerably more compact than those in the local Universe, thereby increasing their resilience to tidal shock-driven disruption and contributing to their survival until the present day.
\end{abstract}

\begin{keywords}
galaxies: formation --- stars: formation -- galaxies: star clusters: general -- globular clusters: general
\end{keywords}


\section{Introduction}
\label{sec:Intro}
Star clusters are subject to a wide variety of dynamical effects that significantly shape their evolution across time. Two-body relaxation leads to a gradual expansion of the cluster half-mass radius \citep[e.g.][]{spitzer_1987, heggie_hut_2003, gieles_etal_2010}. Meanwhile, the most massive objects in the cluster ``sink'' towards the cluster center as two-body relaxation drives the cluster constituents towards equipartition of kinetic energy \citep{spitzer_1969, watters_etal_2000}. The presence of an external tidal field further complicates the evolution, leading to accelerated mass loss from the cluster \citep{henon_1961, spitzer_chevalier_1973, gieles_baumgardt_2008, gieles_etal_2011}. Furthermore, ``tidal shocks''-- impulsive encounters with other massive perturbers, such as passing molecular clouds -- heat the cluster constituents, allowing them to more easily escape beyond the tidal radius and causing the cluster's half-mass radius to change in a manner which depends upon the internal structure of the cluster \citep{gnedin_ostriker_1997, gnedin_ostriker_1999, gnedin_etal_1999a, gnedin_etal_1999b, gieles_renaud_2016, webb_etal_2019}. 

Unfortunately, directly simulating both star cluster formation and the ensuing dynamical evolution remains well out of reach of hydrodynamic galaxy formation simulations (but see \citealt{lahen_etal_2019} and \citealt{ma_etal_2019} for the first steps towards this goal). Therefore, most models typically must adopt computationally cheaper alternate approaches, such as semi-analytic methods of ``painting'' on star cluster formation using (physically-motivated) sub-grid models on top of galaxy formation simulations \citep{pfeffer_etal_2018, choksi_etal_2018, kruijssen_2019_emosaics}. Alternatively, although they do not spatially resolve clusters, \cite{li_etal_sim1, li_etal_sim2} directly model the star cluster formation process in galaxies by self-consistently tracking both the growth of clusters due to accretion from the interstellar medium and the end of formation due to the cluster's own feedback. In all cases, the subsequent dynamical evolution is then accounted for using either prescriptions calibrated against $N$-body simulations or analytic calculations \citep[e.g.][]{pfeffer_etal_2018, choksi_etal_2018,  kruijssen_2019_emosaics}.

Consequently, all such models must make approximations regarding the initial properties of individual clusters. However, the dynamical evolution of a cluster over time depends quite sensitively on the adopted initial conditions. In particular, the initial cluster mass and radius, which together set the initial cluster density, are vital to accurately modeling the mass loss rate from the cluster as essentially all mechanisms for cluster disruption scale with the cluster density \citep[e.g.][]{spitzer_1987,gieles_etal_2006a, gieles_etal_2010, gieles_etal_2011, kruijssen_2011,webb_etal_2019}. Across all quantities needed to describe the initial demographics of a cluster population, most have been studied extensively \citep[such as the fraction of star formation occurring in bound clusters and the resulting initial cluster mass function -- including its lower cutoff, slope, and upper cutoff, e.g.][]{kruijssen_2012_cfe, guszenjov_hopkins_2015, reina-campos_kruijssen_2017, li_etal_sim1, trujillo-gomez_etal_2019}. By contrast, the physics responsible for setting the initial radii of stellar clusters remains poorly understood (but we note that a first attempt was made by \citealt{murray_2009}, who calculated $\Rstar \propto \Mstar^{0.6}$ for very massive clusters and ultracompact dwarf galaxies with $\Mstar \gtrsim 10^{6} \Msun$ whose formation is truncated by radiation pressure).

On the observational side, for clusters within the Local Group where individual member stars can be resolved, the cluster mass can be determined by fitting the colour-magnitude diagram \citep{elson_etal_1989, johnson_etal_2016}. Outside the Local Group, where acquiring detailed CMDs is not possible, photometric mass estimates can be made from mass-to-light ratios calibrated against simple stellar population models or if multi-band photometry is available, from fitting of the spectral energy distribution (SED) \citep[][]{ryon_etal_2015, adamo_etal_2015, ryon_etal_2017}. In other cases, dynamical estimates from velocity dispersions are possible \citep{bastian_etal_2006, mccrady_graham_2007, leroy_etal_2018}. Measuring the radii of stellar clusters is usually more challenging than measuring their masses, as doing so requires the cluster to be at least partially resolved so as to construct a light profile. Radius measurements are further complicated by the fact that only light can be measured, and therefore observations can typically only ever measure a half-light radius, which may differ by a factor of order unity from the half-mass radius due to mass segregation within the cluster \citep{portegies-zwart_etal_2010}. 

Despite the aforementioned challenges, over the past few decades many groups have investigated the properties of young star clusters across a variety of star-forming environments in the nearby Universe \citep[e.g.][]{whitmore_etal_1993, zhang_fall_1999, scheepmaker_etal_2007, mayya_etal_2008, bastian_etal_2012, adamo_etal_2015,   portegies-zwart_etal_2010, bastian_etal_2013, ryon_etal_2015,chandar_etal_2016, ryon_etal_2017, johnson_etal_2017,messa_etal_2018, adamo_etal_2020, adamo_etal_2020b}. The subset of these studies measuring cluster radii generally support an increase in mean cluster radius with cluster \textit{age} (as would be expected from relaxation-driven expansion), but found limited or no evidence in support of a correlation between a cluster's radius and mass (though such conclusions were not universal; see e.g.\ \citealt{bastian_etal_2013}). Consequently, most works simply fit the radius distributions of cluster populations as log-normal distributions of varying means and dispersions.

However, in general all these works were limited by the relatively narrow dynamic range in cluster masses covered by each individual study. Very recently, \cite{krumholz_etal_2019} compiled data on radii and masses for a large sample of young clusters in nearby star-forming galaxies. With the vastly expanded dynamic range, they showed clearly that young star clusters do follow an approximately power-law mass-radius relation $\Rstar \propto \Mstar^{\alpha}$ over the mass range $\Mstar \approx 10^{2} - 10^{7} \Msun$ (see their fig.~9). The heterogeneity of the sample makes determining an exact value of the slope $\alpha$ is uncertain, but it nevertheless appears to be in the range $\alpha \approx 0.25 - 0.33$. The scatter in the relation around the mean trend is roughly 0.4 dex. This large scatter, combined with the relatively weak slope of the underlying MRR, explains why many past individual studies with their more limited dynamic range in cluster mass often concluded against the existence of any MRR. 

Significant strides have been made towards simulating the formation of bound star clusters beginning from the collapse of individual turbulent giant molecular clouds \citep[GMCs; ][]{dale_etal_2015,kim_ostriker_2018, grudic_etal_2018_feedbackfails, howard_etal_2018,li_etal_2019, he_etal_2019} to the final expulsion of the parent cloud gas. These high resolution simulations incorporate a wide variety of relevant physics, including self-consistent modeling of star formation and the subsequent energy and momentum deposition into the surrounding gas. However, they are naturally computationally intensive and therefore cannot sample a large range of parameter space. Moreover, they generally treat the GMCs in isolation, ignoring any impact of the large-scale galactic environment. 

In this work, we take an alternate approach. We seek to develop a simple, physically-motivated analytical model for understanding the initial radii of star clusters across a broad range in mass and as a function of galactic environment. This paper is organized as follows. In Section~\ref{sec:methodology} we present our model for the initial cluster mass-radius relation. Section~\ref{subsec:rsig}--\ref{subsec:starbursts} test the model against observations of young cluster populations across a diverse range of environments in the local Universe. We discuss the implications of our results for a wide range of related problems in Section~\ref{sec:discussion} and conclude with a summary of our main results in Section~\ref{sec:conclusions}. 

\section{Model}
\label{sec:methodology}
\subsection{The initial mass-radius relation}
We begin by considering a single self-gravitating giant molecular cloud (GMC). In particular, we consider an overdense clump within the GMC that ultimately collapses to form a bound star cluster. Throughout, we refer to the initial mass and radius of this clump as $\Mb$ and $\Rb$. In terms of the virial parameter $\avir$, the properties of the clump are \citep{bertoldi_mckee_1992}:
\begin{equation}
    \Rb  = \frac{G\Mb\avir}{5\sigb^2}.
    \label{eqn:vir_blob}
\end{equation}
In reality, there is of course no exact boundary between the clump and the remainder of the cloud, and the central star cluster formed in the central molecular core grows hierarchically through mergers with smaller subclusters which may form closer to the outskirts of the cloud \citep{maschberger_etal_2010,grudic_etal_2018_feedbackfails, grudic_etal_2018_hierarchical,grudic_etal_2020,li_etal_2019}. However, we consider such complications beyond the scope of our simple model.

The mass and radius of the clump are ultimately related to the mass $\Mstar$ and radius $\Rstar$ of the resulting stellar cluster. Not all gas is converted into stars and stellar feedback eventually drives any remaining gas out from the cluster, thereby shallowing the potential in which the remaining stars orbit. Therefore, we relate the clump and stellar cluster properties by applying the simple analytic result derived by \cite{hills_1980}. In the limit that an amount of mass $\Delta M$ is rapidly expelled from a self-gravitating system, the relation between the initial and final radius of the system is  \citep{hills_1980}:
\begin{equation}
    \frac{R}{R_0} = \frac{M_0 - \Delta M}{M_0 - 2\Delta M}
    \label{eqn:hills}
\end{equation}
where $R_0$ and $M_0$ are the initial radius and mass. This result is derived under the assumption that mass loss happens on timescales much shorter than the free-fall time in the clump. Hydrodynamic simulations of collapsing clouds that model both star formation and its feedback effects \citep{grudic_etal_2018_feedbackfails, li_etal_2019}, as well as measurements of the spatial de-correlation length of molecular gas and star formation at small scales in nearby galaxies \citep{kruijssen_heisenberg2, kruijssen_2019_nature, chevance_etal_2020,chevance_etal_2020b,kim_etal_2021}, find that cloud-scale gas expulsion happens on timescales shorter than the cloud free-fall time, and here we assume that this result extends to the smaller scale of the clump. This effectively assumes that the gas expulsion is impulsive. 

The expelled gas mass $\Delta M$ can be expressed in terms of an integrated star formation efficiency in the clump $\epsilon_{\rm C}$, defined as the ratio of the final stellar mass to the initial gas mass. In terms of $\epsilon_{\rm C}$, the results can be succinctly summarized as:
\begin{align}
    \Mstar &= \epsilon_{\rm C} \Mb \nonumber \\ 
    \Rstar &= \left(\frac{\epsilon_{\rm C}}{2\epsilon_{\rm C} - 1}\right)\Rb
    \label{eqn:hills2}
\end{align}
where $\Rstar$ and $\Mstar$ are the resulting stellar cluster radius\footnote{Given the approximate nature of this work, we do not specify $\Rstar$ as e.g.\ a half-mass radius. This slight ambiguity can simply be absorbed into the two order-unity free parameters we describe in this section, $\epsilon_{\rm C}$ and $f_{\rm acc}$.} and mass. In Section~\ref{subsec:params} we discuss reasonable values of $\epsilon_{\rm C}$ in more detail. 

The average velocity dispersion in the clump $\sigb$ can further be related to the average velocity dispersion of the entire cloud $\sigcl$. Application of the size-linewidth relation yields \citep{larson_1981}:
\begin{align}
    \sigb^2 &= \sigcl^2\left(\frac{R_{\rm clump}}{\Rcl}\right) \nonumber  \\ 
    &= \frac{5 \sigcl^4 R_{\rm clump}}{G\Mcl \avir},
    \label{eqn:sigb2}
\end{align}
where the second equality follows from the virial theorem applied to the entire cloud. In reality, the size-linewidth relation in regions of massive star formation is shallower and higher in normalization than the Larson relation adopted here \citep{fuller_myers_1992, caselli_myers_1995, plume_etal_1997}. At least in part, this trend is caused by the well-known result that the normalization of the size-linewidth relation depends on the gas pressure or kpc-scale surface density as $\sigma\propto\sigg^{1/2}$ \citep[e.g.][]{heyer_etal_2009,field_etal_2011,sun_etal_2018,sun_etal_2020}. Below, we account for this trend by adopting an environmentally dependent normalization to the size-linewidth relation\footnote{Stellar feedback may also be partly responsible for the observed variation in the size-linewidth relation \citep{murray_etal_2018}. If so, the shallower relation would be a result of star formation and thus not be applicable at the early stages of star formation/clump collapse relevant in equation \ref{eqn:vir_blob}. }.

Plugging the above expressions into equation~(\ref{eqn:vir_blob}), we obtain:
\begin{align}
\Rstar &= \frac{\avir}{5}\left(\frac{\epsilon_{\rm C}^{1/2}}{2\epsilon_{\rm C} - 1}\right)\frac{ G\Mstar^{1/2}M_{\rm cloud}^{1/2}}{\sigma_{\rm cloud}^2}.
\label{eqn:dummy}
\end{align} 
We set the cloud mass using the model of \cite{reina-campos_kruijssen_2017}. They consider the effects of stellar feedback and centrifugal forces and define the maximum cloud mass as the local \citet{toomre_1964} mass $M_T$, modulo the fraction of the cloud $f_{\rm coll}$ that can collapse under the influence of self-gravity before the cloud is dispersed by feedback:
\begin{align}
    M_{\rm cloud} &= f_{\rm coll}M_T \nonumber \\ 
    &= f_{\rm coll}\frac{\sigma_g^4}{G^2\sigg Q^4} .
    \label{eqn:Mcloud}
\end{align}
Here, $Q$ is Toomre's stability parameter \citep{toomre_1964}, $\sigma_g$ and $\sigg$ are the velocity dispersion and total gas surface density in the ISM (averaged over kpc scales). \cite{reina-campos_kruijssen_2017} and \citet{pfeffer_etal_2019} show that this model provides a good match to observations of molecular clouds in a wide range of environments.

Building on \cite{krumholz_mckee_2005} and \cite{kruijssen_2012_cfe}, we assume that star and cluster formation proceed within a galactic disc in hydrostatic equilibrium, allowing us to relate $\sigcl$ to the gas velocity dispersion in the ISM $\sigma_g$ as: 
\begin{equation}
    \frac{\sigma_{\rm g}}{\sigcl}  = f_{\rm coll}^{-1/4}\left(\frac{\phi_{\rho}}{\phiPbar}\right)^{1/2},
    \label{eqn:sigs}
\end{equation}
where $\phi_{\rho}$ and $\phiPbar$ are order unity constants representing respectively the ratio of the mean cloud density and pressure to the corresponding disc midplane properties. The factor of $f_{\rm coll}^{-1/4}$ in Equation~\ref{eqn:sigs} ensures that the size-linewidth relation is maintained as the collapse fraction varies (i.e.\ $M_{\rm cloud}/\sigcl^4$ is independent of $f_{\rm coll}$). Equations \ref{eqn:Mcloud} and \ref{eqn:sigs} imply that $\sigcl$ is higher in regions where $\Sigma_{\rm g}$ is higher, in line with the higher normalization of the size-linewidth relation in regions of high ISM pressure or intense star formation \citep{sun_etal_2018,sun_etal_2020}.
\cite{krumholz_mckee_2005} further show that:
\begin{equation}
\label{eq:phiP}
\phi_{\bar{P}} \approx 10 - 8f_{\rm GMC},
\end{equation}
where
\begin{equation}
\label{eqn:fgmc}
    f_{\rm GMC} = \frac{\Sigma_{{\rm H}_2}}{\Sigma_{\rm g}}\approx\left[1+0.025\left(\frac{\sigg}{10^2~\Msun~\pc^{-2}}\right)^{-2}\right]^{-1},
\end{equation}
is the fraction of all gas locked in GMCs (assuming all molecular gas is in GMCs). For a disc in hydrostatic equilibrium, they also show that $\phi_{\rho}$ can be written as:
\begin{equation}
    \phi_{\rho} = \left(\frac{375}{2\pi^2}\right)^{1/4}\left(\frac{\phiPbar^3}{\phi_P\alpha_{\rm vir}^3}\right)^{1/4},
    \label{eqn:phi_rho}
\end{equation}
where $\phi_P$ is an order-unity correction factor to account for the fact that the gravity of the stars compresses the gas. From the Galactic disc to starbursting environments, \cite{krumholz_mckee_2005} find $\phi_{P} \approx 3$, and therefore we adopt this as a constant value throughout our calculations.

A final additional complication is that accretion onto proto-stars during the cluster formation process causes stars to slow down and fall deeper into the potential, shrinking the cluster's size \citep{bonnell_etal_1998, bonnell_clarke_2008, moeckel_clarke_2011, kruijssen_etal_2012}. We parameterize the magnitude of this effect in an order-unity correction factor $f_{\rm acc} \lesssim 1$ to $\Rstar$. Though in principle the factor $f_{\rm acc}$ could depend on the mass of the cluster \citep{gieles_etal_2018}, for simplicity in this work we assume $f_{\rm acc} = 0.6$, as measured in the simulations of \citet{bonnell_etal_2008} by \cite{kruijssen_etal_2012}.

Combining Equations \ref{eqn:Mcloud}-\ref{eqn:phi_rho} we obtain the mass-radius relation:
\begin{align}
\Rstar &= \left(\frac{3}{10\pi^2}\frac{\avir}{\phi_{P}\phi_{\bar{P}}}\right)^{1/4}\frac{\epsilon_{\rm C}^{1/2}}{2\epsilon_{\rm C} - 1}\frac{f_{\rm acc}}{Q^2}\left(\frac{\Mstar}{\sigg}\right)^{1/2}     \label{eqn:rm}  \\
&\approx 1.5\,\mathrm{pc}\left(\frac{\Mstar}{10^4\Msun}\right)^{1/2}\left(\frac{\sigg}{10\Msun \mathrm{pc}^{-2}}\right)^{-1/2} \nonumber 
\end{align}
where in calculating a fiducial normalization for $\Rstar$ we adopt $\phi_P=3$, $f_{\rm GMC} = 1$, $\alpha_{\rm vir} = 2.0$, $\epsilon_{\rm C} = 1.0$, $f_{\rm acc}=0.6$, and $Q=2$. We discuss the expected range of these parameters in more detail in Section \ref{subsec:params}.

The fiducial form of equation~(\ref{eqn:rm}) is remarkably similar to the clump mass-radius relation observed in the Milky Way by \citet{urquhart_etal_2014}, which follows $\Rb/\pc\approx2.2~(\Mb/10^4~\Msun)^{0.6}$. This close correspondence to our cluster mass-radius relation is encouraging, given that our model assumes a close correspondence between the clump radius and the cluster radius (see equation~\ref{eqn:hills2}). The uncertainties on the model parameters are such that these relations are entirely consistent, especially when varying the accretion-induced shrinkage ($f_{\rm acc}\neq0.6$) or adding feedback-driven expansion ($\epsilon_{\rm C}<1$).

All quantities appearing before $(\Mstar/\sigg)^{1/2}$ in equation~(\ref{eqn:rm}) are dimensionless and order-unity. Thus, our model predicts that a characteristic size for stellar clusters is imprinted by the the surface density of gas in the galactic disc $\sigg$. This result therefore predicts an environmentally dependent initial cluster (surface) density. We also emphasize that our model makes no statement about whether clusters of a given $\Mstar$ will form. Rather, it only states that if they form, their radii should approximately obey equation~(\ref{eqn:rm}). In reality, cluster formation at certain mass regimes may be suppressed due to the local environment \citep{kruijssen_2012_cfe, adamo_etal_2015, reina-campos_kruijssen_2017, li_etal_sim1, trujillo-gomez_etal_2019}. We return to this point in Section~\ref{subsec:shallowing_slope}.

\subsection{Dynamical evolution}
\label{subsec:dyn_evolution}
The definition of what constitutes a ``young'' stellar cluster in the literature varies wildly, with ``young'' clusters in the literature ranging from $\sim1$~Myr to $\sim1$~Gyr \citep[e.g.][]{krumholz_etal_2019}. Meanwhile, dynamical evolution within the cluster proceeds on a relaxation timescale \citep{spitzer_1987}:
\begin{equation}
    t_{\rm rlx} = 40\,\mathrm{Myr} \left(\frac{N}{10^4}\right)^{1/2}\left(\frac{\Rstar}{1 \rm pc}\right)^{3/2}\left(\frac{\bar{m}}{1 \Msun}\right)^{-1/2}\left(\frac{\ln\Lambda}{5}\right)^{-1}, 
    \label{eqn:trlx}
\end{equation}
where $N$ is the total number of stars in the cluster, $\bar{m}$ is their mean mass, and $\ln\Lambda$ is the Coulomb logarithm, where $\Lambda \approx 0.02N$ for clusters with a standard stellar mass function \citep{giersz_heggie_1994a}. If $\Rstar \propto \Mstar^{\alpha}$, then $t_{\rm rlx}$ increases with cluster mass for all $\alpha > -1/3$. Plugging in our MRR derived in the previous section, $\Rstar \propto \Mstar^{1/2}$, we obtain $t_{\rm rlx} \propto M^{5/4}$ (assuming $\bar{m}$ is constant so that $N \propto \Mstar$). Therefore,  low-mass clusters go through their dynamical evolution faster than high-mass clusters. For low-mass clusters with $\Mstar \lesssim 10^4 \Msun$, equation~(\ref{eqn:trlx}) shows that the relaxation time may be much less than the typical cluster age. Therefore, two-body relaxation cannot be ignored and internal dynamical evolution should have an important role in shaping the radii of low-mass, young clusters.

\cite{gieles_etal_2010} used $N-$body simulations of isolated clusters in the absence of a tidal field to investigate the evolution of the cluster radius with time. They found that in dynamically young ($t \ll t_{\rm rlx}$) clusters, mass-loss due to stellar evolution provides an energy source that leads to a mild expansion of the cluster, similar to the effect of relaxation. In the opposite limit of dynamically old clusters, relaxation dominates the cluster evolution and continues to drive expansion. They combine these two limiting cases continuously in the following expression for the evolution of the cluster radius and show it provides an excellent fit to their numerical results:
\begin{equation}
    \Rstar(t) = \Rstar(t=0)\left( \left[\frac{t}{t_{\star}}\right]^{0.14} + \left[\frac{\chi(t) t}{t_{\rm rlx}(t=0)}\right]^{4/3} \right).
    \label{eqn:rh_ev}
\end{equation}
The dimensionless parameter $\chi(t)$ measures the width of the stellar mass function in the cluster, which evolves with time due to ejection of low-mass stars from the cluster. \cite{gieles_etal_2010} find that it evolves approximately as:
\begin{equation}
    \chi(t)= 3\left(\frac{t}{t_{\star}}\right)^{-0.3},
    \label{eqn:chi_ev}
\end{equation}
where $t_{\star} \equiv \mathrm{min(2\,Myr}, t)$.

This prescription implicitly neglects the regulating effect of the local tidal field on the dynamical evolution \citep[e.g.][]{gieles_etal_2011}. However, for most young clusters, we expect the instantaneous local tidal field to have a subdominant effect on the radius evolution \citep[e.g.][]{ryon_etal_2017, krumholz_etal_2019, rui_etal_2019}. Nevertheless, to prevent unreasonable levels of expansion we approximately take into account the effect of the tidal field by imposing a maximum cluster radius. \cite{alexander_etal_2014} find that the ratio of the cluster's half-mass to tidal radius
\begin{equation}
    f_{\rm roche} \equiv \left(\frac{\Rstar}{R_{\rm tid}}\right)_{\rm max},
    \label{eqn:f_roche}
\end{equation}
does not exceed a value of approximately 0.3. The tidal radius $R_{\rm tid}$ can be written in terms of the galactocentric angular velocity $\Omega$ at the cluster's position as:
\begin{equation}
    R_{\rm tid} = \left(\frac{G\Mstar}{2\Omega^2}\right)^{1/3}.
\end{equation}
We apply this result as an upper limit to the allowed cluster radius, i.e.\ if our calculated $\Rstar$ ever exceeds $f_{\rm roche}R_{\rm tid}$ we set $\Rstar = f_{\rm roche}R_{\rm tid}$. 

This upper limit only truncates the radii of clusters forming in extreme environments with especially strong tidal fields (e.g.\ the Central Molecular Zone of the Milky Way, see Section~\ref{subsec:starbursts}). Apart from such exceptional cases, and because $f_{\rm roche}$ only enters into our model as an upper limit to the cluster radius, we find that our results are independent of its value for $f_{\rm roche} \gtrsim 0.2$.

\subsection{Exploration of the parameter space and parameter choices} 
\label{subsec:params}

\begin{figure*} 
\includegraphics[width=\textwidth]{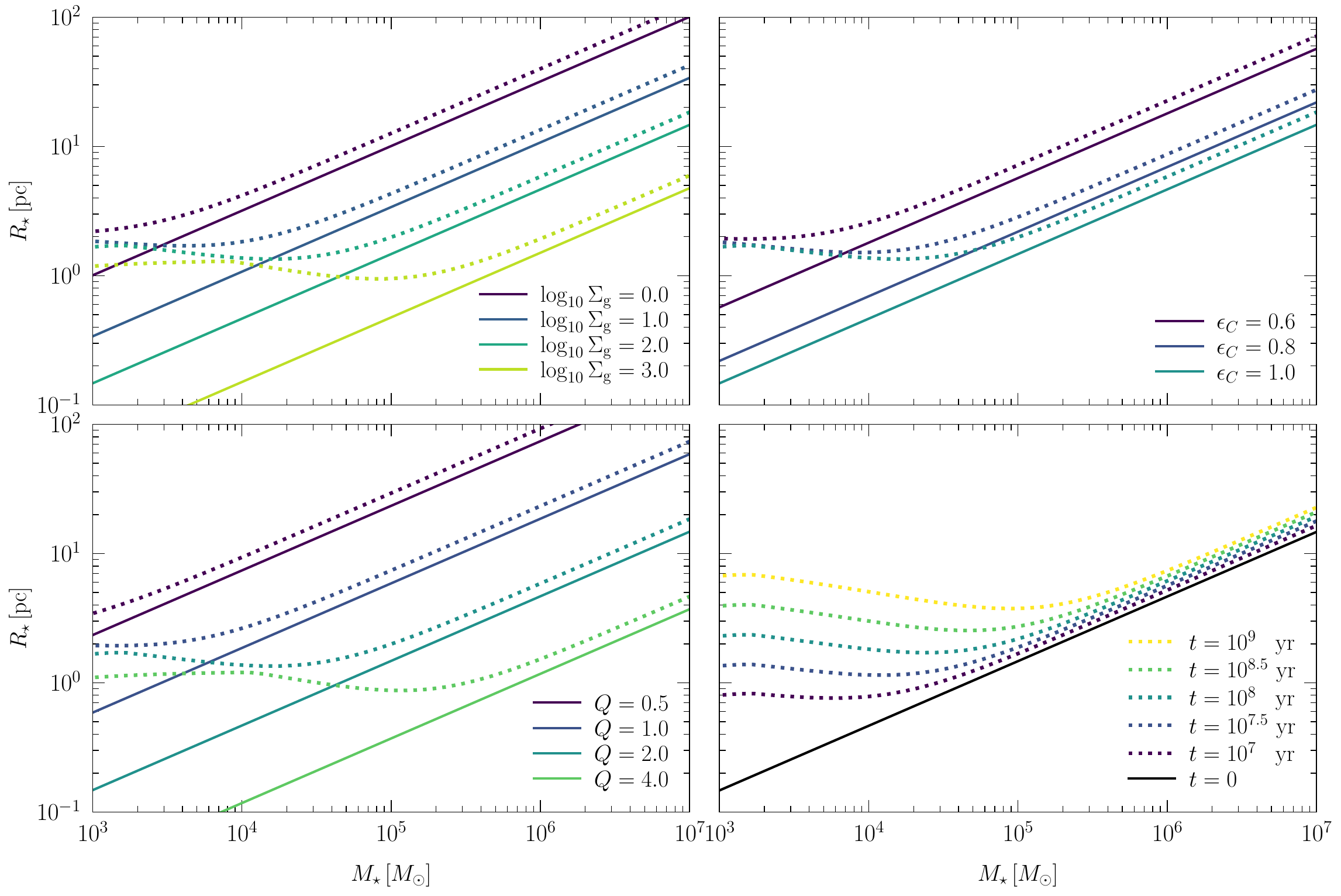}
\vspace{-2mm}
\caption{Impact of varying different model parameters. \textit{Top-left:} total gas surface density $\sigg$ (units of $\Msun~\pc^{-2}$). \textit{Top-right:} integrated star formation efficiency $\epsilon_{\rm C}$. \textit{Bottom-left:} Toomre $Q$ parameter. \textit{Bottom-right:} cluster age $t$. Solid lines show the initial, dynamically unevolved relation [equation~(\ref{eqn:rm})] and dotted lines show the result of applying equation~(\ref{eqn:rh_ev}) to dynamically evolve the solid curves by two-body relaxation. For the purposes of this figure, we adopt fiducial parameters  $\sigg = 10^2~\Msun~\pc^{-2}$, $\epsilon_{\rm C} = 1$, $Q = 2.0$, $t = 50$~Myr, except in the panel in which the given parameter is varied. }
  \label{fig:param_space}
\end{figure*}

The top-left panel of Fig.~\ref{fig:param_space} shows the effect of varying the most important parameter in our model, the galactic gas surface density $\sigg$. The adopted range of $\sigg$ is chosen to cover essentially the entire dynamic range of the Kennicutt-Schmidt relation, from the outskirts of local dwarf galaxies where $\sigg \sim 10^{0.5} \Msun~\pc^{-2}$ to starbursting galaxies at high-redshift with $\sigg \sim 10^{3.5}\, \Msun~\pc^{-2}$ \citep[e.g.][]{shi_etal_2018}. Over this range, the expected variation in $\Rstar$ is quite significant (a factor of about 20) and corresponds to variations in initial cluster densities of $\sim 10^4$. We note, however, that such a large range in density at fixed cluster mass is unlikely to be realised in nature, simply because the formation of massive, bound stellar clusters requires high surface densities to begin with \citep{kruijssen_2012_cfe, reina-campos_kruijssen_2017, li_etal_sim1}. We return to this point in more detail in Section~\ref{subsec:shallowing_slope}.

The bottom left panel of Fig.~\ref{fig:param_space} shows the variation of $\Rstar$ with the Toomre $Q$ parameter. We test a range of $Q$ spanning the range found in typical star forming regions, and find that the major impact of disc stability on the degree of gravitational collapse affects the initial cluster radius correspondingly. More stable discs produce lower-mass clouds that host smaller clumps and therefore produce smaller clusters. Across a plausible range of Toomre $Q$ parameters, we find a variation of the initial cluster radius by nearly two orders of magnitude.

The top right panel of Fig.~\ref{fig:param_space} shows the effect of varying the integrated star formation efficiency for $\epsilon_{\rm C} \geq 0.6$. The higher values of $\epsilon_{\rm C}$ lead to more compact clusters by mitigating the effects of gas expulsion from the cluster. The adopted range of $\epsilon_{\rm C}$ leads to a factor of $\sim10$ variation in cluster radii. In the context of our model $\epsilon_{\rm C}$ refers to the integrated star formation efficiency \textit{within the clump}, which is assumed to be an overdense region within the GMC that ultimately forms the bound stellar cluster. Consequently, the conversion of gas into stars in the clump can be significantly higher than the average over the entire cloud (which is observed to be a few per cent, see \citealt{kruijssen_2019_nature} and \citealt{chevance_etal_2020,chevance_etal_2020c}), because many local free-fall times can elapse within the clump before feedback sets in and expels gas (see also Section~\ref{subsec:discussion_limits} for further discussion on this point). The value of $\epsilon_{\rm C} = 1$ is slightly higher than typical values of $\epsilon_{\rm C} \approx0.5$ taken in past similar works \citep[e.g.][]{matzner_mckee_2000, kruijssen_2012_cfe}. 

The dotted lines in all of these first three panels demonstrate the effect of allowing the initial MRR to evolve due to two-body relaxation for $t = 50$ Myr. Clusters that have masses corresponding to $t_{\rm rlx} \lesssim t$, i.e.\ clusters that are dynamically old, are clearly identifiable as the MRR develops a break to the left of the corresponding critical mass scale. The dynamically evolved clusters quickly grow to significantly larger radii than set by our initial MRR. Indeed, these panels demonstrate that relaxation washes away memory of the initial MRR and the initial parameter choices in the limit that $t \gg t_{\rm rlx}$, as the dotted curves begin to pile up regardless of their initial radii. However, for dynamically young (i.e.\ at fixed age more massive) clusters with $t < t_{\rm rlx}$ our predicted initial MRR remains essentially unchanged, except for a slight expansion due to energy-input driven by stellar evolution, which of course proceeds on timescales of a few Myr irrespective of the cluster properties.

In the bottom-right panel of Fig.~\ref{fig:param_space} we consider the effects of varying the cluster age. The age range shown of $10^6{-}10^9$ yr brackets essentially the widest possible range of ``young'' cluster definitions in the literature, but we caution that our adopted prescription for two-body relaxation is likely an insufficient descriptor of the cluster dynamics for $t \gg 100$ Myr, as the external tidal field eventually has a non-negligible effect on the cluster evolution. The adopted spread in ages leads at most to a factor of ten variation in cluster radius at $\Mstar \sim 10^3 \Msun$, but this spread decreases rapidly with increasing cluster mass, as relaxation times quickly increase too. 

The accretion-induced shrinkage effect (for brevity we do not vary $f_{\rm acc}$ in Fig.~\ref{fig:param_space}) provides a mild adjustment to the normalization of the cluster MRR. \cite{kruijssen_etal_2012} found in the hydrodynamic simulations of collapsing clouds by \citet{bonnell_etal_2008} that this effect leads to about a factor of about 1.6 decrease in cluster radii (corresponding to $f_{\rm acc} \approx 0.6$), as measured from roughly the start of star formation until the first cloud free-fall time had elapsed. The value they measure is only approximate, given the arbitrary time range over which this measurement is carried out. Given this uncertainty on its exact value, we choose to simply adopt a fiducial value of $f_{\rm acc}=0.6$ and find that this matches the data well. Finally, we note that the exact values of $\epsilon_{\rm C}$ and $f_{\rm acc}$ are degenerate: for a constant overall normalization to $\Rstar$ in equation~(\ref{eqn:rm}), lower values of $\epsilon_{\rm C}$ in turn permit lower values of $f_{\rm acc}$ (and vice-versa) and so we have chosen a combination of values that lies close to independent constraints on each parameter.

\subsection{Population synthesis}
\label{subsec:popsynth}
Observed cluster populations show a significant amount of scatter in their MRR \citep{krumholz_etal_2019}. Therefore, before delving into a comparison of our model and observed cluster populations (Section~\ref{sec:results}) we first outline here a simple population synthesis model to generate a mock cluster population.

We begin by setting the virial parameter $\avir$ of the cloud from which the cluster formed. Using data from the PHANGS-ALMA survey \citep{leroy_etal_2020,leroy_etal_2020b}, \cite{sun_etal_2018} measured the virial parameters of molecular gas in a sample of 15 nearby galaxies. They find typical values of $\avir \approx 2$ with a dispersion of about $\sigma_{\avir} \approx 0.2$ dex and we adopt these values as the mean and dispersion for a log-normal distribution from which to draw $\avir$. As discussed in Section~\ref{subsec:params}, the exact values of $\avir$ make little difference in the MRR.

Having set the parent cloud properties, we next draw clusters randomly in galactocentric radius $\Rgc$. Given a radial gas surface density profile (e.g.\ from observations), we sample $\Rgc$ in proportion to the area-weighted gas surface density, i.e.
\begin{equation}
    \frac{dN}{d\Rgc} \propto \Rgc \Sigma_g(\Rgc),
\end{equation}
where $\Sigma_g$ is the mean gas surface density at a galactocentric radius $\Rgc$. We also account for azimuthal scatter in $\sigg$ at fixed $\Rgc$. In their sample of 33 nearby spiral galaxies, \cite{schruba_etal_2011} measure typical azimuthal variations of $\sigma_{\Sigma_g} \approx 0.3$ dex and so we adopt this value for our model. We show in Section~\ref{sec:results} that this azimuthal scatter plays a significant role in setting the overall scatter in the MRR.

Once a cluster's $\Rgc$ and $\sigg$ are set as described above, we also assign a cluster mass. We draw cluster masses from an initial cluster mass function (ICMF) of the form:
\begin{equation}
    \frac{dN}{d\Mstar} \propto \Mstar^{-2}\,\,\mathrm{for}\,\,10^3 \Msun < \Mstar < \Mc.
\end{equation}
This straightforward functional form is motivated by numerous observations of young cluster populations in the nearby universe \citep[e.g.][]{krumholz_etal_2019}. In reality, detailed analysis reveals deviation from pure power-law behaviour at high masses, with the ICMF functional form better described by a \cite{schechter_1976} function $dN/d\Mstar \propto e^{-\Mstar/\Mc}$ \citep{gieles_etal_2006a, larsen2009, bastian_2008, portegies-zwart_etal_2010, adamo_etal_2015, johnson_etal_2017, messa_etal_2018}. For simplicity, we neglect this correction here and simply truncate the ICMF at $\Mc$.

We choose to limit ourselves to $\Mstar > 10^3~\Msun$, because below that mass the identification of truly bound clusters as opposed to unbound associations is extremely difficult \citep[e.g.][]{krumholz_etal_2019}. Moreover, our model implicitly assumes that the resulting stellar cluster forms as the dominant structure within its parent GMC. Therefore, our model may be less accurate at describing lower mass clusters. 

The upper cutoff mass $\Mc$ of the ICMF has now been robustly demonstrated both theoretically \citep{li_etal_sim1, reina-campos_kruijssen_2017, pfeffer_etal_2019} and empirically \citep{jordan_etal_2007, johnson_etal_2017} to vary strongly with galactic environment.\footnote{We note that \cite{trujillo-gomez_etal_2019} have recently proposed that the \textit{minimum} mass varies with galactic environment as well. We have neglected this possibility here, given that we only consider clusters with $\Mstar>10^3~\Msun$.} In particular, such studies generally found that the mass function extends to higher masses as the local gas or star formation rate surface density increase. We account for this effect using the analytic model presented in \cite{reina-campos_kruijssen_2017}. In brief, their model sets the maximum stellar cluster mass as the product:
\begin{equation}
M_c(Q, \sigg, \Omega) = \epsilon_{\rm cl} \Gamma M_{\rm cl, max},
\end{equation}
where $Q$ is the \cite{toomre_1964} stability parameter, $\epsilon_{\rm cl}$ is the integrated star formation efficiency of the entire parent GMC, $\Gamma$ is the bound cluster formation efficiency \citep{bastian_2008}, and $M_{\rm cl, max}$ is the maximum cloud mass. They set $\Gamma$ using the analytic cluster formation efficiency model of \cite{kruijssen_2012_cfe}, which has been shown to match observational constraints in the local Universe \citep{adamo_etal_2020} and numerical simulations \citep{li_etal_sim2,grudic_etal_2020}. Finally, they assume the maximum cloud mass is shear-limited (i.e.\ the Toomre mass), except in regions where stellar feedback would destroy the cloud before collapse of the shear limited region could complete. They show this model reproduces observed masses of clusters, clouds, and star-forming clumps both in the local Universe and at high-redshift. This environmentally varying truncation of the ICMF plays an important role in shaping the MRR, as we discuss in the next subsection.

\begin{figure*} 
\includegraphics[width=\textwidth]{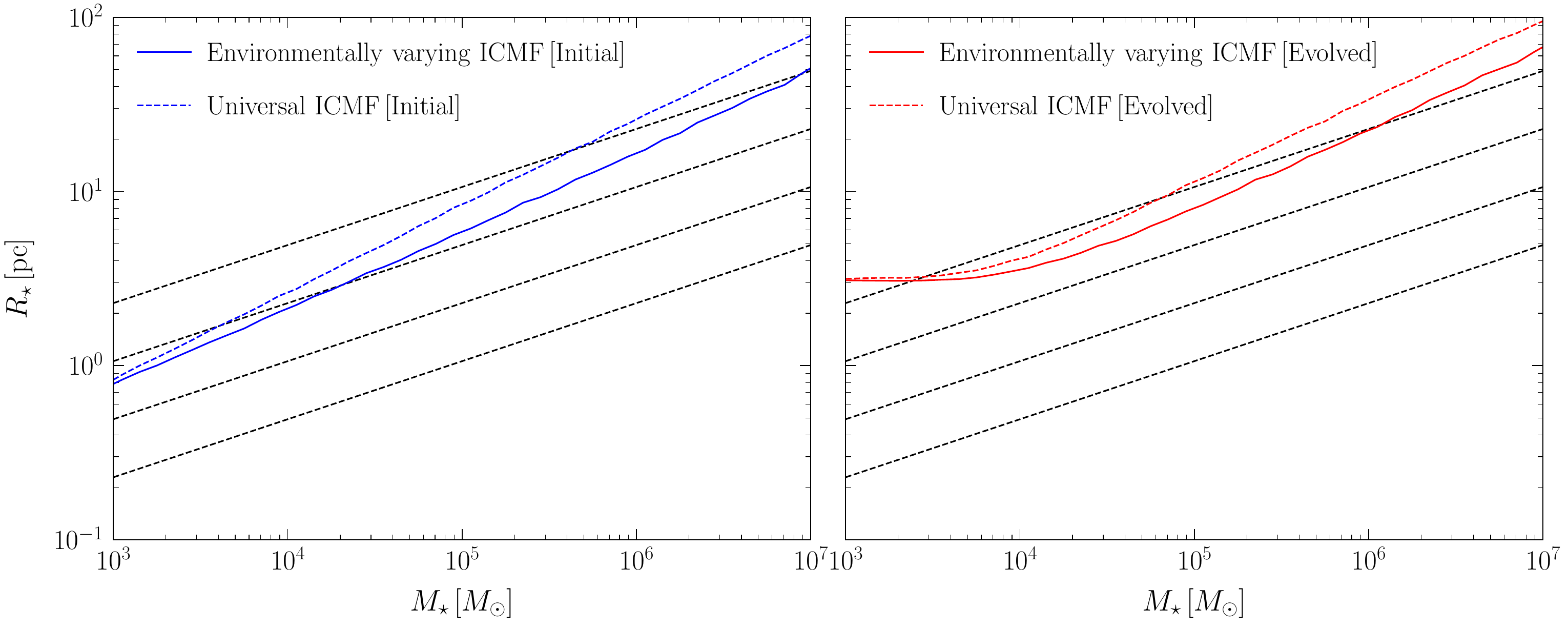}
\vspace{-2mm}
\caption{\textit{Left:} Predicted initial MRR for a mock cluster population, in which we draw clusters from both a universal initial cluster mass function (ICMF; dotted curve) and an environmentally dependent ICMF (solid curve). Note that we have applied the same population synthesis model as in Section \ref{subsec:popsynth} and adopted a mock galaxy profile as described in Section \ref{subsec:shallowing_slope}. \textit{Right:} Same as left panel, but now showing the median MRR after dynamical evolution. For reference, the black dashed lines mark constant volume densities ($Rstar\propto\Mstar^{1/3}$). Enforcing a more realistic, environmentally varying, ICMF causes massive clusters to preferentially form in high gas-density environments, lowering the normalization of their MRRs and therefore shallowing the \textit{global} MRR slope (see text for further discussion).}
  \label{fig:slope_test}
\end{figure*}

\subsection{How an environmentally varying maximum cluster mass reshapes the MRR}
\label{subsec:shallowing_slope}
Our prediction for the initial mass-radius relation in equation~(\ref{eqn:rm}) is $\Rstar \propto \Mstar^{1/2}$. And yet, in the compilation of clusters from \cite{krumholz_etal_2019}, the global slope appears to be somewhat shallower. While directly fitting the \cite{krumholz_etal_2019} sample is unreasonable given the large biases and arbitrary choices involved in weighting different populations in a fit, the scaling in their compilation nevertheless appears closer to $\Rstar \propto \Mstar^{1/3}$. We argue that the resolution to these two discrepant scalings is two-fold.

As we have already demonstrated (e.g.\ Fig.~\ref{fig:param_space}), at the low-mass end of the MRR two-body relaxation washes out any memory of the initial MRR and quickly shallows the slope of the MRR (e.g.\ Fig.~\ref{fig:param_space}). In the limit $t \gg t_{\rm rlx}$, two-body relaxation actually drives clusters towards an inverse MRR $\Rstar \propto \Mstar^{-1/3}$ \citep{gieles_etal_2010}. However, for typical ``young,'' low-mass clusters with $t \sim t_{\rm rlx}$, the expected scaling is quite shallow and lies somewhere in between $\Rstar \propto \Mstar^{-1/3}$ and $\Rstar \propto \Mstar^{1/2}$, as is clearly visible in the figures throughout this paper.

In addition, we argue the MRR slope shallows at the high-mass end, due to the combined effect of the environmental dependence of the initial cluster mass function (ICMF) and the initial cluster mass-radius relation. Because high-mass clusters can \textit{only} form in high-density environments, they cannot populate curves in mass-radius space corresponding to $\sigg$ less than some critical $\Sigma_{g, \rm crit}$. Because $\Rstar \propto \Mstar^{1/2}\sigg^{-1/2}$, this increase in the typical $\sigg$ of high mass clusters offsets the increase in radius due solely to their mass. In other words, the \textit{global} slope -- i.e.\ the slope that one would obtain when fitting composite populations across many orders of magnitude in mass -- is the result of ``piecing together'' many $\Rstar \propto \Mstar^{1/2}$ curves, but with decreasing normalization as one moves rightward in the MRR plane. As a result, the global slope decreases.

We can obtain a rough estimate of this effect as follows. Using a sample of local galaxies and their young cluster populations, \cite{johnson_etal_2017} provide an empirical fit between $\Mc$ and $\Sigma_{\rm SFR}$:
\begin{equation}
    \log_{10}\frac{\Mc}{\Msun}= 1.07\log_{10}\left(\frac{\Sigma_{\rm SFR}}{\Msun\,\, \rm yr^{-1}\,\,kpc^{-2}}\right) + 6.82.
    \label{eqn:mc_sfr}
\end{equation}
We can relate $\Sigma_{\rm SFR}$ to $\sigg$ via the Kennicutt-Schmidt relation \citep{kennicutt_1998}:
\begin{equation}
    \frac{\Sigma_{\rm SFR}}{\Msun\,\,\mathrm{yr^{-1}\,\,kpc^{-2}}} = 2.5 \times 10^{-4} \left(\frac{\sigg}{\Msun \,\, \rm pc^{2}}\right)^{1.4}.
    \label{eqn:sfr_sigg}
\end{equation}
Combining these scaling relations and assuming clusters of mass $\Mc$ form at the minimum gas density required to form them, we obtain $\Rstar \propto \Mstar^{0.17}$, notably shallower than our initial $\Rstar \propto \Mstar^{1/2}$. We emphasize that the slope of 0.17 we derive here is only a very rough estimate meant to demonstrate how the environmental dependencies of the cluster radius and mass function combine to shallow the global MRR slope. In reality, the \cite{johnson_etal_2017} relation is only an approximation valid on galaxy-averaged scales \citep[see also][]{pfeffer_etal_2019} and breaks down on smaller scales, which is why we choose instead to use the theoretical model of \cite{reina-campos_kruijssen_2017}.

We illustrate more precisely the combination of aforementioned effects by applying the population synthesis model as described in the previous subsection for both an environmentally dependent power-law ICMF and a universal ICMF with fixed $\Mc = 10^7 \Msun$. For simplicity, we construct a mock galaxy with surface density profile $\sigg(\Rgc) \propto \Rgc^{-1}$, normalized to $\sigg = 100 \Msun\,\rm pc^{-2}$ at 1 kpc. We take a flat, Milky Way-like rotation curve with
$\Omega = V_{\rm rot}/R_{\rm GC}$, with $V_{\rm rot} = 220$~km~s$^{-1}$ and $Q \approx 1$ across the galaxy. 

The left and right panels of Fig.~\ref{fig:slope_test} show the median initial MRR for the case of a universal (dotted) and environmentally varying (solid) ICMF; the latter clearly tends to shallows the global slope. We see in the right panel of Fig.~\ref{fig:slope_test} that the low-mass end of the MRR also becomes shallower after the clusters undergo dynamical evolution, and the slope over the entire range from $10^3 \Msun$ to $10^7 \Msun$ ends up tracking more closely the dashed lines of constant density, $\Rstar \propto \Mstar^{1/3}$.

\begin{figure} 
\includegraphics[width=\columnwidth]{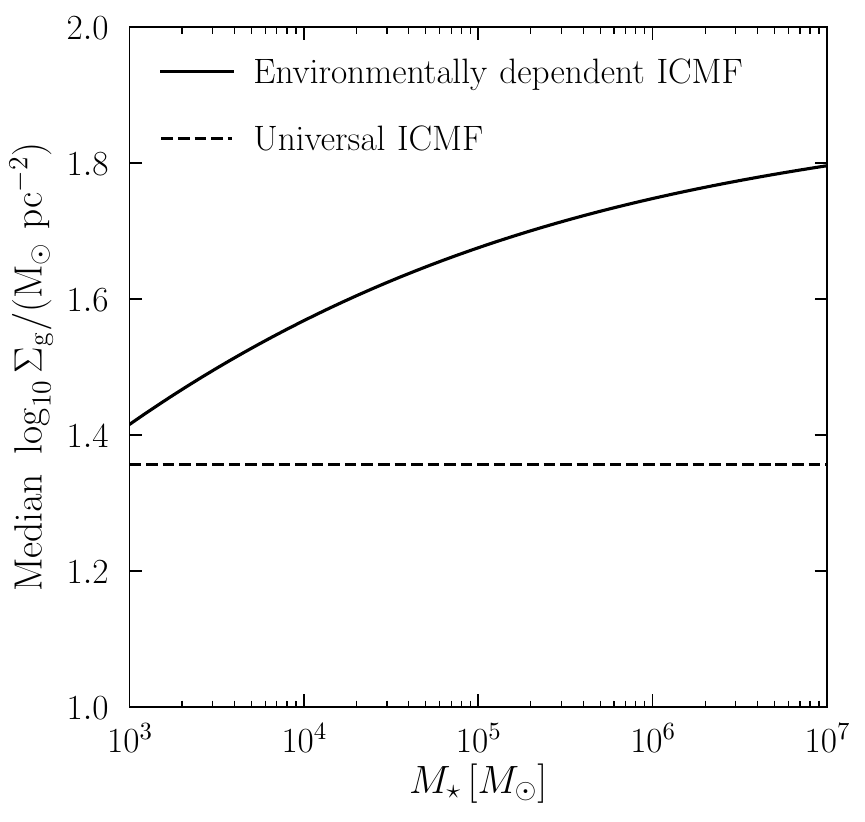}
\vspace{-2mm}
\caption{Median gas surface density $\sigg$ at formation as a function of cluster mass for the mock cluster population plotted in Fig.~\ref{fig:slope_test}. The solid curve shows that massive clusters preferentially form in higher gas density environments, lowering the normalization of their MRR [equation~(\ref{eqn:rm})] and thereby shallowing the global MRR slope. For a universal ICMF (dashed line), on the other hand, the median $\sigg$ does not vary with cluster mass because clusters of any mass are allowed to form at any surface density.}
  \label{fig:sig_mass}
\end{figure}

In Fig.~\ref{fig:sig_mass} we also show how the median gas surface density $\overline{\Sigma}_g$ at formation changes as a function of cluster mass for the mock cluster population shown in Fig.~\ref{fig:slope_test}. For the case of a universal ICMF (dashed line), the median gas surface density is of course simply constant, because clusters do not know about the large-scale environment. On the other hand, for an  environmentally varying ICMF (solid line) the median gas surface density increases towards higher cluster masses.

Given the large scatter in the observed relation, we argue that the intrinsic form we propose here is consistent with observations. We also emphasize that Fig.~\ref{fig:slope_test} is only intended to illustrate the general effect of an environmentally varying ICMF on the MRR. The precise slope and normalization of the global MRR will ultimately depend on the details of how cluster populations are sampled across different galaxies as well as the galactic surface density, rotation, and stability parameter profiles. Finally, we note that the effect we have demonstrated here is not unique to any particular scaling relation between $\Rstar$ and $\Mstar$; so long as higher density environments produce more compact clusters and the maximum cluster mass increases with density too, the global MRR slope must be shallower than the local one.

\section{Comparison to observed cluster populations}

\subsection{Testing the environmental dependence}
\label{subsec:rsig}
A fundamental prediction of our model is that the initial radii of stellar clusters vary with galactic environment. In particular, equation~(\ref{eqn:rm}) predicts that high gas surface density environments form more compact clusters. The existence of such a relation has not yet been systematically examined in the literature (though hints of it were noticed by \cite{bastian_etal_2013}, who showed that typical cluster radius increased with galactocentric radius in NGC7252, and \citealt{kruijssen_2015}, who noted that clusters are more compact in the high-density environment of the Galactic Centre than in galaxy discs). Therefore, we begin by testing this model prediction using a sample of cluster populations in six different galaxies spanning a wide range of environments (note that we defer a complete description of the adopted data sources to Sections~\ref{subsec:discs} and~\ref{subsec:starbursts}). 
Given the limited dynamic range in cluster radius and $\sigg$ in any single galaxy, we choose not to look for such a scaling within the cluster populations of individual galaxies, but instead to compare the typical compactness of clusters across different galaxies. 

To measure stellar cluster compactness, we perform a $\chi^2$ minimization fit for the overall normalization $k$ of the galaxy's stellar cluster population's MRR while enforcing a functional form $\Rstar = k \Mstar^{1/2}$. To mitigate the effects of dynamical evolution, we fit only clusters with masses $M > 10^4 \Msun$. Using the best-fit parameters, we then calculate the median $\Rstar(\Mstar = 10^{4.5} \Msun)$ in that galaxy.

\begin{figure}
\includegraphics[width=\columnwidth]{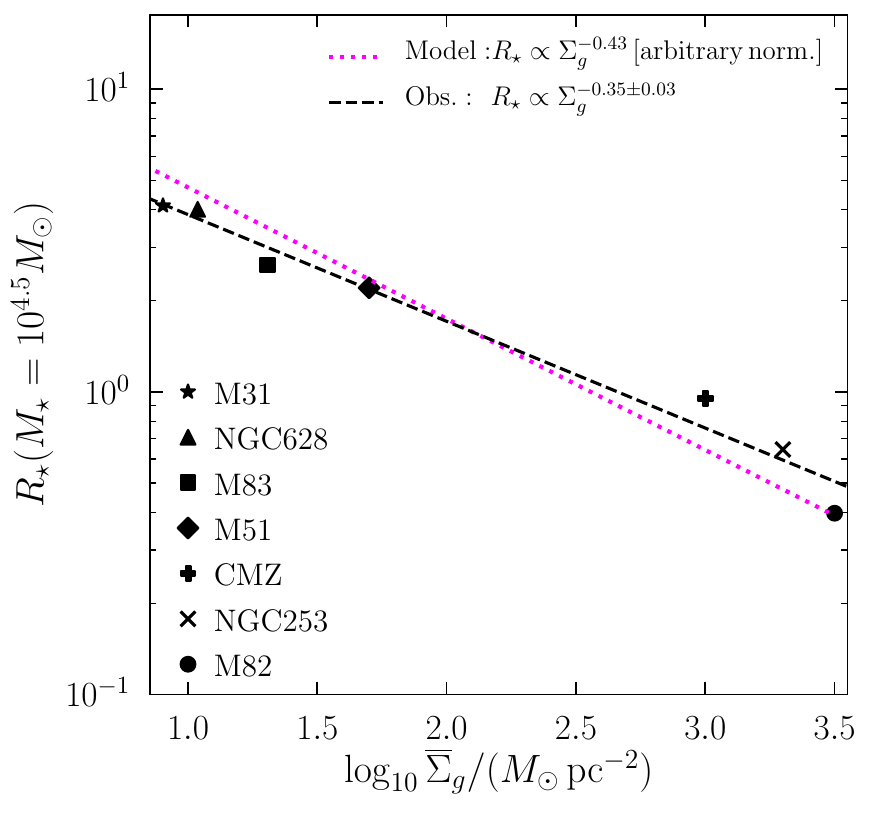}
\vspace{-5mm}
\caption{Demonstration of environmental dependence of initial cluster radii. The $x$-axis gives the (area-averaged) gas surface density of the galaxy and the $y$-axis gives the typical cluster radius at $\Mstar = 10^{4.5} \Msun$. The seven data points represent different observed systems (M82, NGC 253, NGC 628, M51, M31, M83, and the Milky Way's CMZ) and their corresponding cluster populations. The best-fit power-law relation describing the data is represented by the black dashed line, which scales as $\Rstar \propto \sigg^{-0.35}$. The magenta dotted curve shows a power-law fit to the model prediction (scaled to an overall normalization constant). Note that the model dependence of $\Rstar$ on $\sigg$ is somewhat weaker than $\propto \sigg^{-0.5}$ [as in equation~(\ref{eqn:rm})] because we have allowed $f_{\rm GMC}$ to scale with $\sigg$ [see equation~(\ref{eqn:fgmc}) and the accompanying discussion]. }
  \label{fig:rsig_obs}
\end{figure}

As a simple indicator of the overall galactic environment we use the area-averaged gas surface density $\overline{\Sigma}_g$:
\begin{equation}
\overline{\Sigma}_g = \frac{\int_0^{R_{\rm GC, max}} 2\pi \Rgc \sigg(\Rgc) d\Rgc}{\pi R_{\rm GC, max}^2}.
\label{eqn:avg_sigg}
\end{equation}
Fig.~\ref{fig:rsig_obs} shows the resulting relationship between the normalization of the MRR and the averaged gas surface density. The best-fit power-law relation for the observed cluster populations is $\Rstar \propto \sigg^{-0.35}$, in qualitative agreement with the expected scaling $\Rstar \propto \sigg^{-0.5}$ predicted in equation~(\ref{eqn:rm}). The agreement between the model and observed relations is even better when we take into account the fact that the molecular gas fraction $f_{\rm GMC}$ (on which our predicted MRR has a weak dependence) also varies with $\sigg$, as given by equation~(\ref{eqn:fgmc}). This dependence on $\sigg$ yields a mild deviation from the naive scaling between $\Rstar$ and $\sigg$, so that the best-fit power-law relation describing the two quantities is $\Rstar \propto \sigg^{-0.43}$, bringing the model and observed scalings between cluster radius and environment into good quantitative agreement (see the magenta dotted curve in Fig.~\ref{fig:rsig_obs}).

\label{sec:results}

\subsection{Nearby disc galaxies: M31, M51, M83, NGC628}
\label{subsec:discs}
We begin by testing our model against cluster populations in several typical star-forming disc galaxies. Our model requires the total gas surface density $\sigg$ and the molecular gas fraction $f_{\rm GMC}$ to calculate $\Rstar(\Mstar)$. Furthermore, to calculate the maximum cluster mass as a function of environment -- relevant for drawing clusters from the ICMF during population synthesis -- we additionally require the Toomre stability parameter and angular velocity profiles $Q(\Rgc)$ and $\Omega(\Rgc)$. We compile all the relevant data from the observational literature for the galaxies and cluster populations against which we compare. For all galaxies, we compile both atomic and molecular gas surface density profiles independently. Therefore, we can calculate the molecular gas fraction $f_{\rm GMC} = \Sigma_{H_2}/(\Sigma_{H_2} + \Sigma_{HI})$ directly from the data rather than from the approximate relation given in equation~(\ref{eqn:fgmc}). We note that the radial range covered by the observed profiles generally does not encompass the entire galaxy. Rather than perform uncertain extrapolation of the relevant profiles, we simply limit ourselves to drawing clusters within the observed range. Below we describe the relevant sources of data used for comparison. Readers that are only interested in the comparison between the model and observed populations can skip to Section~\ref{subsubsec:discs_results}.

\subsubsection{M31}
We use the cluster catalogue of \cite{johnson_etal_2015}, with the original data taken as part of the Panchromatic Hubble Andromeda Treasury Survey \citep[PHAT; ][]{dalcanton_etal_2012}. At the close distance of M31, it is possible to resolve individual stars in clusters. Therefore, cluster masses were calculated in \cite{johnson_etal_2016} by fitting the observed colour-magnitude diagram using the MATCH code of \cite{dolphin_2002}, while cluster effective radii $R_{\rm eff}$ were calculated in \cite{johnson_etal_2015} from the measured light profile in band F475W and interpolating to determine the radius enclosing half the light. We take the molecular and atomic gas surface density profiles from \cite{schruba_etal_2019}, with the initial data described in \cite{caldu-primo_etal_2016}. As part of the Westerbork Synthesis Radio Telescope HI survey of M31, \cite{braun_etal_2009} calculated the velocity dispersion profile of atomic gas and \cite{corbelli_etal_2010} calculated a galactic rotation curve. From these data, we calculate the Toomre stability parameter $Q$ as a function of galactocentric radius:
\begin{equation}
    Q = \frac{\kappa\sigma_g}{G \sigg},
    \label{eqn:toomre}
\end{equation}
where $\sigma_g$ is the measured velocity dispersion of the atomic gas (justified by the low molecular gas fraction in M31) and $\kappa$ is the epicyclic frequency. For simplicity, we approximate $\kappa = \sqrt{2}\Omega$. 

\subsubsection{M51}
We use the properties of M51 stellar clusters calculated by \cite{chandar_etal_2016}. They calculate cluster masses by fitting the observed broadband SEDs and estimate $R_{\rm eff}$ values from the light profile using the \textsc{IShape} software \citep{larsen_1999}. \cite{schinnerer_etal_2013} and \cite{colombo_etal_2014} characterized the molecular gas distribution in M51 as part of the Plateau de Bure Interferometer Arcsecond Whirlpool Survey (PAWS). From these data, \cite{lang_etal_2020} calculated rotation curves and $Q$ profiles  as part of the PHANGS-ALMA survey \citep{leroy_etal_2020}. Finally, we take the atomic gas profiles from \cite{schuster_etal_2007}.

\subsubsection{M83}
We take the properties of stellar clusters in M83 from \cite{ryon_etal_2015}. They calculated $R_{\rm eff}$ for their clusters by applying \textsc{galfit} \citep{peng_etal_2002} and calculated cluster masses from fitting the observed SEDs following the method of \cite{adamo_etal_2010}. We note that roughly 20\% of clusters in the \cite{ryon_etal_2015} catalogue have radii $> 100$ pc. Not only are these radii associated with very large error-bars (as discussed in their sect.~4.1), but we also consider them unlikely to represent truly bound structures and so we discard them for the purposes of comparison to our model. For the host galaxy, we take atomic gas surface density profiles and rotation curves from \cite{walter_etal_2008}. We calculate $Q$ from these data as well as the molecular gas surface density and velocity dispersion profiles (justified by the high molecular gas fraction in M83) reported in \cite{freeman_etal_2017}.

\subsubsection{M74}
The final disc galaxy we consider is M74 (NGC 628). We take properties of stellar clusters from \cite{ryon_etal_2017}, with data obtained as part of the Legacy Extragalactic UV Survey (LEGUS). They again calculated cluster radii using \textsc{galfit} applied to the observed light profiles and calculated masses from fitting the observed SED (using the fitting code \textsc{yggdrasil} of \citealt{zackrisson_etal_2011}). We again use galactic rotation curves and $Q$ profiles derived by \cite{lang_etal_2020} as part of the PHANGS-ALMA survey \citep{leroy_etal_2020}. Finally, we take the atomic gas profiles calculated by \cite{walter_etal_2008} as part of the THINGS survey and tabulated by \cite{schruba_etal_2011}.

\begin{figure*} 
\includegraphics[width=\textwidth]{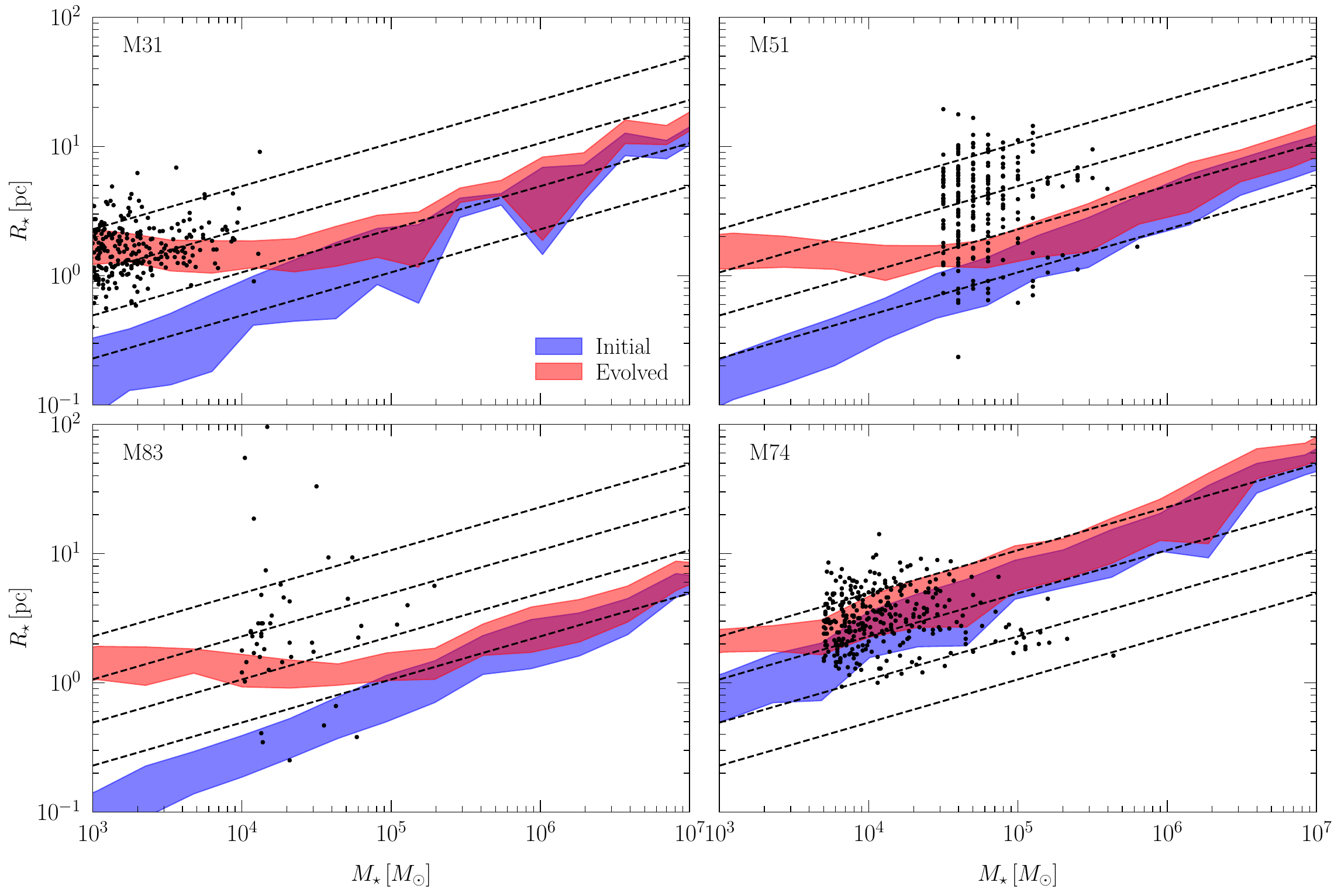}
\vspace{-2mm}
\caption{MRR in four typical star-forming disc galaxies. Blue shows our prediction for the \textit{initial} MRR [equation~(\ref{eqn:rm})] and red shows the dynamically evolved population. Shaded regions show the $\pm 1\sigma$ range for the model cluster population (see text for details on our adopted population synthesis model). Dashed lines mark constant densities of $\rho = 10^1-10^4 \Msun~{\rm pc}^{-3}$, in steps of factors of ten. }
  \label{fig:discs}
\end{figure*}

\subsubsection{Model results in M31, M51, M83, and M74}
\label{subsubsec:discs_results}
Fig.~\ref{fig:discs} shows the result of applying our model to these systems. The blue shaded regions show the initial MRR, as predicted by our equation~(\ref{eqn:rm}), while the red shaded regions show the result of dynamically evolving the model cluster population. Throughout Fig.~\ref{fig:discs} we have adopted a constant cluster formation history and drawn clusters uniformly over the range 10-100 Myr (clusters with ages less than $\sim10$~Myr are omitted to avoid any effects of extinction during the embedded phase, see e.g.\ \citealt{krumholz_etal_2019}). Where age estimates for observed clusters are available, we have applied the same age cut to observed populations. 

In all four galaxies, the model provides a reasonable match to the observed cluster populations. Throughout, it predicts cluster densities of $\rho_{\star} \sim 10^1 - 10^3 \Msun \,\rm pc^{-3}$, similar to the observed range of cluster densities in these galaxies (also see fig.~9 of \citealt{krumholz_etal_2019}). Nevertheless, the evolved model clusters are systematically slightly larger than the observed clusters by 0.1-0.2 dex. In Section~\ref{sec:discussion} we discuss how this normalization offset may be attributable to our omission of cluster shrinkage induced by tidal shocks. Furthermore, though the model provides a simple explanation for the observed global $\Rstar \propto \Mstar^{1/3}$ relation (Section \autoref{subsec:shallowing_slope}), the slopes of the cluster MRR in these four individual galaxies appear to be shallower than that predicted by our model. Some of this discrepancy may be due to the large observational error and intrinsic scatter affecting the cluster MRR. The scatter in the model relation stems from three sources: scatter in the initial cloud virial parameter, azimuthal scatter in the gas surface density, and scatter in cluster ages (though the latter affects only the ``evolved'' curves). Overall, the model scatter (defined by the width of the shaded regions, which encompass the 16th-84th percentiles) is of a similar magnitude as the observed scatter. The median vertical $1\sigma$ scatter in the model MRR is 0.12 dex. For comparison, the median scatter in the observed population in these four galaxies is 0.26 dex. However, part of the observed scatter is also undoubtedly attributable to observational error in measuring cluster sizes. For example, \cite{ryon_etal_2015, ryon_etal_2017} quote typical errors of 0.1-0.2 dex. Therefore, the model scatter of 0.12 dex may be consistent with the true scatter in the relation.

\begin{figure*} 
\includegraphics[width=\textwidth]{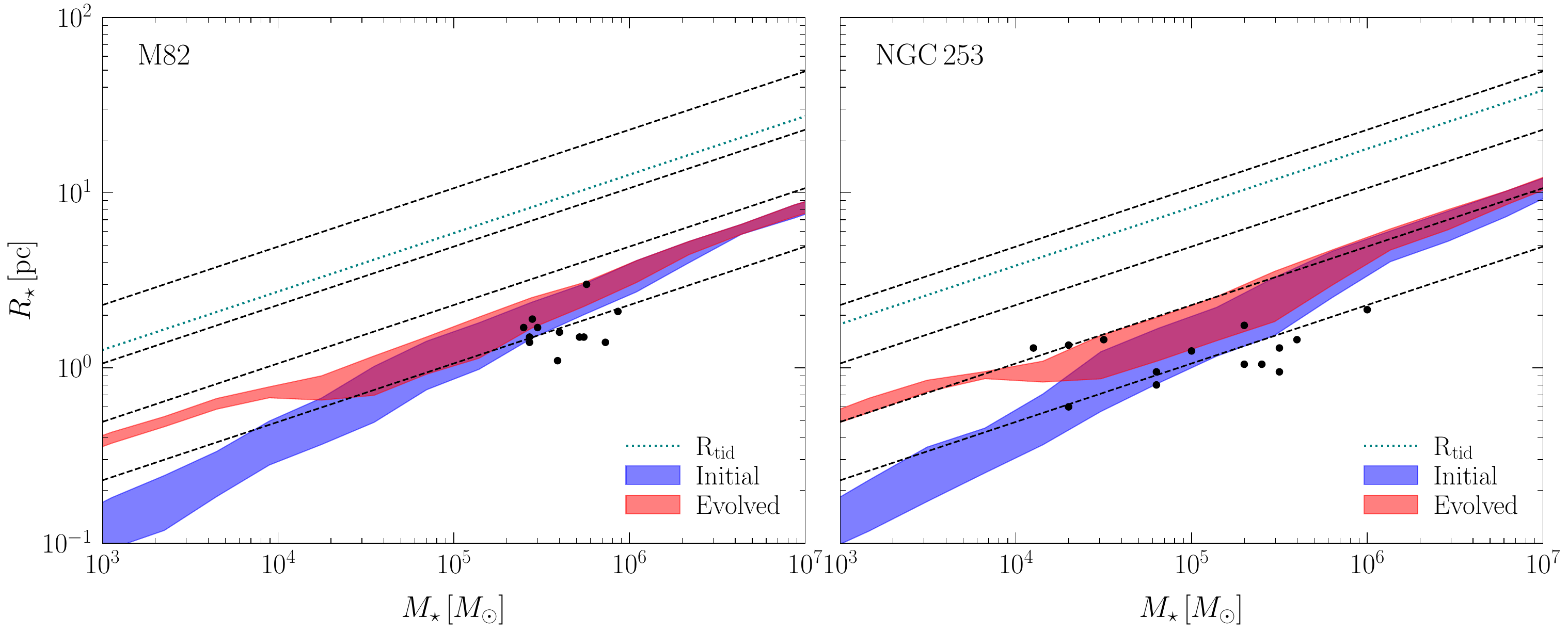}
\vspace{-2mm}
\caption{As in Fig.~\ref{fig:discs}, but here we show the MRR for the two circumnuclear cluster populations in M82 and NGC253. The observed clusters in these galaxies have systematically smaller radii due to the higher gas surface densities in these environments [equation~(\ref{eqn:rm})]. Because we take $\Omega$ as a constant for these regions, we also present the tidal radius as the single cyan dotted line. The model curve at the very lowest masses in M82 (where no observed clusters lie) is affected by the tidal field -- there, the relaxation times are short enough to drive cluster expansion until reaching the adopted upper limit on the cluster radius of $f_{\rm roche}R_{\rm tid}$ (see text for details). }
  \label{fig:starbursts}
\end{figure*}

\subsection{Nuclear starbursts: M82, NGC253, and the Milky Way CMZ}
\label{subsec:starbursts}

We now move on to test our model in three extreme environments: the nuclear starbursting regions of M82 and NGC 253, as well as the central 500 pc of the Milky Way, i.e.\ the Central Molecular Zone (CMZ). 
We adopt constant gas surface densities of $\sigg = 10^{3.5}, 10^{3.3},$ and $10^3 \Msun~{\rm pc}^{-2}$ for these three regions respectively \citep{kennicutt_1998, longmore_etal_2013} and calculate $f_{\rm GMC}$ according to equation~(\ref{eqn:fgmc}) in all three cases. \cite{kennicutt_1998} present orbital periods $t_{\rm dyn}$ at the outer edge of the nuclear regions for M82 and NGC 253. We convert to orbital angular frequency $\Omega = 2\pi/t_{\rm dyn}$, which yields values of $1.0\times10^3$ and $0.62\times10^3$ km $\rm s^{-1}\,kpc^{-1}$ respectively and adopt a value of $\Omega = 1.7\times10^3$ km $\rm s^{-1}\,kpc^{-1}$ for the CMZ \citep{kruijssen_2014_cmz}. 

We take properties of the stellar clusters in  M82 and NGC 253 from \cite{mccrady_graham_2007} and \cite{leroy_etal_2018} respectively. The circumnuclear stellar cluster population of M82 is relatively young ($\sim10$~Myr) and coeval \citep{satyapal_etal_1997, mccrady_graham_2007}, so here we draw cluster ages uniformly over the range 5-15 Myr. Meanwhile, the clusters of NGC 253 are estimated to be even younger, $\sim1$~Myr \citep{leroy_etal_2018}, so here we draw uniformly over the age range 0-5 Myr.

The CMZ contains only two young massive clusters, the Arches and the Quintuplet. We adopt masses of $1.2 \times 10^4 \Msun$ and $2 \times 10^4 \Msun$ and radii of $0.5$ pc and $1.4$ pc for the Arches and Quintuplet, respectively \citep{clarkson_etal_2012, rui_etal_2019}. Both clusters are expected to be roughly co-eval with ages of a few Myr \citep{schneider_etal_2013}, and so we draw ages of model clusters uniformly over the range 0-5 Myr, as in NGC 253.

Fig.~\ref{fig:starbursts} show the predicted MRR of clusters in the extragalactic nuclear environments. These environments host both relatively massive and young clusters, so dynamical evolution is essentially irrelevant. As a result, both the initial and evolved MRRs provide an excellent match to the data. However, in NGC 253, three clusters with masses of $\sim2\times 10^4 \Msun$ lie above the predicted MRR by a factor of $\sim2$. While age estimates of individual clusters are unavailable, one possible explanation for the existence of such clusters is that they represent a slightly older subpopulation -- we find that an evolved model curve corresponding to $\sim20$~Myr provides an excellent match to those three points. Note also that the extremely high gas surface densities in both environments push the maximum cluster mass to $\sim 10^7 \Msun$, and as a result the MRR in both environments does not shallow below $\Rstar \propto \Mstar^{0.5}$. 

\begin{figure} 
\includegraphics[width=\columnwidth]{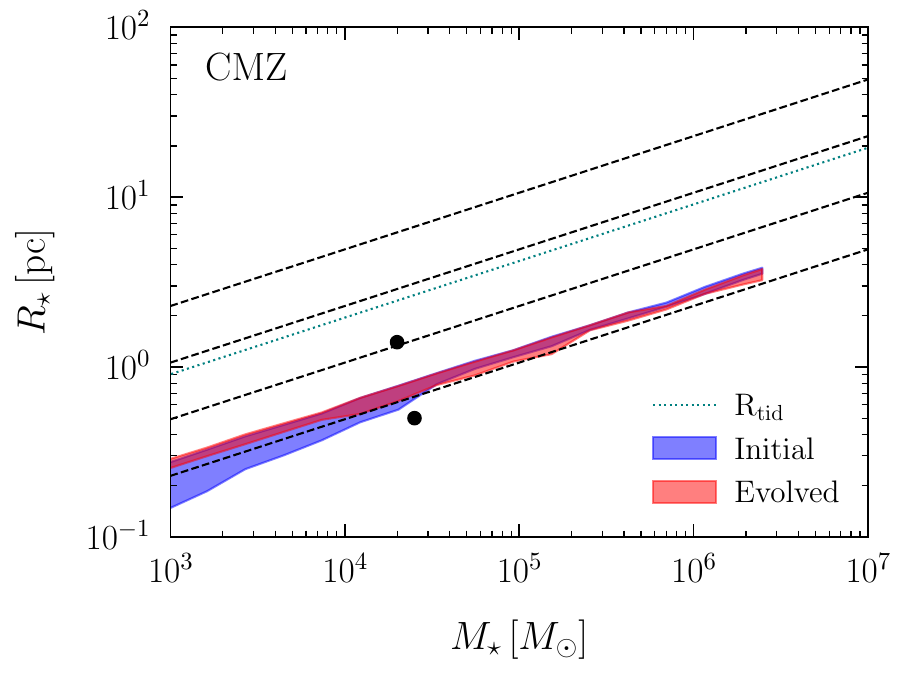}
\vspace{-2mm}
\caption{As in Fig.~\ref{fig:starbursts}, but now for the Milky Way's CMZ. We show the two young massive clusters residing in the CMZ, the Arches and Quintuplet. In the CMZ, the ``initial'' and ``evolved'' regions overlap entirely because of the extremely strong tidal field, which caps the cluster radius at a value $f_{\rm roche}R_{\rm tid}$.  }
  \label{fig:cmz}
\end{figure}

Finally, Fig.~\ref{fig:cmz} shows the predicted MRR in the CMZ. Given the limited number of data points in the CMZ, a detailed comparison is difficult. Nevertheless, the model curve runs reasonably in between the Arches and Quintuplet. It appears to provide a better match for the Arches, which directly overlaps with the model curve. On the other hand, the Quintuplet is a factor of 3 larger than the Arches and is offset from the upper edge of the model contour by roughly 0.3 dex. This offset is largely due to the imposed cap in cluster radius $\Rstar = f_{\rm roche}R_{\rm tid}$ which defines the upper limit of the model contours in the CMZ. Interestingly, \cite{rui_etal_2019} recently found no evidence of tidal truncation in the Quintuplet's structure, hinting that our adopted value of $f_{\rm roche}$ may either be too low or not universal. Alternatively, this may imply that the Quintuplet cluster is on the verge of tidal destruction.

\section{Discussion}
\label{sec:discussion}

\subsection{Implications for globular cluster formation at high-redshift}
\label{subsec:discussion_gcformation}
The past decade has seen significant strides in the modeling of star cluster formation across cosmic time. Many models have demonstrated that the existence of both young stellar clusters and globular clusters can be explained simultaneously and simply as the by-product of the regular star-formation process \citep{muratov_gnedin_2010, kruijssen_2012_cfe, li_gnedin_2014,  kruijssen_2015, li_etal_sim1, choksi_etal_2018, pfeffer_etal_2018, choksi_gnedin_2019a, choksi_gnedin_2019b, el-badry_etal_2019, kruijssen_2019_emosaics, pfeffer_etal_2019, reina-campos_etal_2019, keller_etal_2020}, without the need to invoke new or ``exotic'' physical processes \citep[e.g.][]{peebles_dicke_1968,peebles_1984}. These models have generally successfully matched observed properties of local star cluster populations, including their ages, mass functions, metallicity distribution functions, and the correlations of all these observables with host galaxy properties. The dependence of cluster radius on galactic gas surface density that we present in this work contributes yet another facet of star cluster formation inextricably tied to the build-up of the host galaxy. 

Globular clusters (GCs) formed in high-redshift galaxies with typical gas surface densities considerably higher than those of most young clusters host galaxies today. Higher surface density environments also lead to a higher probability of clusters being tidally shocked and thereby dissolved, the so-called ``cruel cradle'' effect \citep{kruijssen_etal_2012}. However, our model predicts that these higher surface density environments also lead to smaller initial GC radii, and therefore higher initial cluster densities and correspondingly slower GC disruption. This effect may therefore help to counteract the cruel cradle effect and increase the probability of GC survival.

Current models of globular cluster formation typically apply simple prescriptions for the cluster radii. For example,the E-MOSAICS simulations \citep{pfeffer_etal_2018} adopt a constant cluster radius of 4~pc, independent of cluster mass, while the semi-analytic model of \cite{choksi_etal_2018} assumes an initial MRR calibrated against the Galactic globular cluster system, $\Rstar \propto \Mstar^{1/3}$ with a normalization of 2.4 pc at $2\times 10^5 \Msun$. Our model for the MRR can be used to improve the accuracy of such theoretical works and better model the dynamical evolution and destruction of globular clusters across cosmic time. Alternatively, it can be used to motivate regions of the mass-radius parameter space to explore in detailed $N-$body simulations of individual clusters that are computationally limited to studying only a small range of initial conditions.

Cosmological simulations of galaxy and globular cluster formation suggest that a typical value for $\sigg$ at the sites and epochs of globular cluster formation is $\sim 10^{2.5} \Msun~{\rm pc}^{-2}$ \citep{kruijssen_2019_emosaics,keller_etal_2020}, which yields $\Rstar \gtrsim 4.5$ pc for $\Mstar \gtrsim 10^5 \Msun$, though we emphasize that the normalisation of the cluster MRR in our model is uncertain.

Our model can be adapted for a subgrid implementation in hydrodynamic galaxy formation simulations. We rewrite Eq. \ref{eqn:rm} in terms of the local volumetric density $\rho_{\rm g}$ using Eqns. 33-34 of \cite{krumholz_mckee_2005}:
\begin{equation}
    \Rstar = \left(\frac{\pi}{1250}\right)^{1/4}\left(\frac{f_{\rm acc}}{\epsilon_{\rm C}^{3/2}}\frac{\phi_{\rho}\phi_P^{1/4}}{\phiPbar}\right)\left(\frac{\Mstar G^{1/2}}{\sigma_g \rho_g^{1/2} }\right)^{1/2}.
\end{equation}
We expect this expression will be included in future simulations using the MOSAICS model \citep{kruijssen_2011,pfeffer_etal_2018} for star cluster formation and evolution. We also note that while we have adopted a constant $\phi_P \approx 3$ in our calculations, in reality $\phi_P$ varies with galactic environment \citep{pfeffer_etal_2018}, and is generally lower in the outskirts of galaxies where the gas surface density is lower; this spatial variation of $\phi_P$ can be included in simulations \citep[e.g.,][]{pfeffer_etal_2018}, and would further flatten the cluster MRR because $\Rstar \propto \phi_P^{-1/4}$. 

\subsection{Implications for observed proto-GC candidates} 
\label{subsec:discussion_obsprotogc}
Recent observations in gravitationally-lensed HST fields have begun to reveal a population of low-luminosity, relatively compact ($\lesssim$ 50 pc), actively star-forming sources at $z > 2$ \citep{vanzella_etal_2017, bouwens_etal_2017, vanzella_etal_2019}. The inferred masses and star formation rates of the sources appear to be consistent with proto-globular clusters (GCs), though no proto-GC with a radius similar to that of GCs in the local Universe ($\sim3$~pc) has been definitively resolved to date. Our results allow us to predict the expected radii of such possibly unresolved proto-GCs.

\cite{vanzella_etal_2019} report on a proto-GC candidate source at $z \approx 6$ located at the center of a dwarf galaxy. This proto-GC candidate has a star formation rate surface density $\Sigma_{\rm SFR} > 10^{2.5} \Msun\,\rm yr^{-1}\,kpc^{-2}$. The radius of the source is constrained using a \textsc{galfit} \citep{peng_etal_2002} model to be $\lesssim 13$ pc, while SED modeling yields a mass estimate of $\Mstar \lesssim 10^6 \Msun$. 

Zooming out, they find that the host of this proto-GC source is a dwarf galaxy with radius $R\approx450$~pc. The host's $\Sigma_{\rm SFR}$ is somewhat lower, but still significant, $\Sigma_{\rm SFR} \approx 10^{1.4} \Msun\,\rm yr^{-1}\,kpc^{-2}$. We assume the host galaxy obeys the Kennicutt-Schmidt relation measured by \cite{genzel_etal_2010} for star-forming galaxies at $z \sim 2$:
\begin{equation}
    \frac{\Sigma_{\rm SFR}}{\Msun\,\,\mathrm{yr^{-1}\,\,kpc^{-2}}} = 3.3 \times 10^{-4} \left(\frac{\sigg}{\Msun \,\, \rm pc^{2}}\right)^{1.17}.
    \label{eqn:sfr_sigg2}
\end{equation}
Application of this expression yields a galactic gas surface density $\sigg \approx 10^{4.17} \Msun/\rm pc^2$. Applying our MRR in equation~(\ref{eqn:rm}), we predict $\Rstar \approx 2.7$ pc, about a factor of 5 smaller than their quoted upper limit. We note that in reality the Kennicutt-Schmidt relation may not be applicable to this galaxy, e.g.\ due to a lack of sufficient number of star-forming clumps in the galaxy over which to average \citep{kruijssen_heisenberg1, kruijssen_heisenberg2}. However, we find that $\sigg$ would still have to decrease by a factor of $\gtrsim$20 before the cluster radius grew to the upper limit of 13 pc placed by \cite{vanzella_etal_2019}. 

Nevertheless, it seems likely that most GC formation in the Universe happened in less extreme environments than that of the \cite{vanzella_etal_2019} source, as discussed in Section~\ref{subsec:discussion_gcformation}. Our model predicts that a $10^6 \Msun$ cluster forming in an environment with $\sigg \approx 500 \Msun/\rm pc^2$ would have an initial radius of 14 pc. Therefore, it may be possible for future observations to achieve sufficient spatial resolution to marginally resolve individual clusters \citep[see also ][]{renzini_2017,zick_etal_2018, pfeffer_etal_2019}.

\subsection{Implications for globular clusters as hosts of LIGO sources}
\label{subsec:discussion_ligo}
Stellar clusters have received renewed interest for their possible roles as factories of dynamically assembled black hole binaries observable by LIGO \citep[e.g.][]{portegies-zwart_mcmillan_2000, morscher_etal_2015, rodriguez_etal_2015, rodriguez_etal_2018, antonini_etal_2019}. However, recent work has highlighted how the predicted merger rates depend sensitively on the initial cluster radii. \citep[][]{hong_etal_2018, rodriguez_loeb_2018, choksi_etal_2019, antonini_gieles_2019}. A more complete understanding of the initial cluster MRR is therefore vital for isolating the precise mechanism(s) responsible for driving LIGO sources to coalescence. Our model, and others like it, can provide a physically motivated framework guide future numerical studies of binary black hole production in clusters.

\subsection{Comparison to numerical simulations}
\label{subsec:sims_comparison}
Complementary to our simple analytical model, several recent numerical simulations have investigated the nature of the cluster mass-radius relation. \citet{smilgys_bonnell_2017} have modelled star cluster formation in hydrodynamical simulations of molecular clouds that are perturbed by a spiral shock. They obtain an approximately linear mass-radius relation, with $R_\star/\pc\approx1.0(M_\star/10^4~\Msun)$. This steep slope is inconsistent with both our model and with observations, which is likely caused by the omission of stellar-dynamical expansion and an environmentally-dependent maximum cloud mass in the simulations, whereas both of these are included and flatten the mass-radius relation in our model.

\citet{grudic_etal_2020} provide a more recent numerical experiment aiming to reproduce the observed cluster mass-radius relation. These authors perform a comprehensive suite of isolated cloud simulations in which the initial cloud properties are varied systematically, resulting in the mass-radius relation:
\begin{equation}
    \label{eq:grudic}
    \frac{R_\star}{\pc}\approx3.0~\left(\frac{M_\star}{10^4~\Msun}\right)^{\frac{1}{3}}\left(\frac{\Sigma_{\rm cloud}}{10^2~\Msun~\pc^{-2}}\right)^{-1}\left(\frac{M_{\rm cloud}}{10^6~\Msun}\right)^{\frac{1}{5}}\left(\frac{Z}{{\rm Z}_\odot}\right)^{\frac{1}{10}} ,
\end{equation}
where $Z$ is the metallicity. The normalisation and cluster mass slope closely match our analytical results, but the dependence on cloud surface density is considerably steeper than our predicted dependence on the kpc-scale gas surface density ($R_\star\propto\sigg^{-1/2}$), even if the sign of the dependence is the same. In order for both relations to be consistent, the cloud surface density would need to increase as $\Sigma_{\rm cloud}\propto\sigg^2$. However, the cloud surface density contrast relative to the large-scale interstellar medium decreases towards larger surface densities and pressures \citep[e.g.][]{kruijssen_2015}, so that the discrepancy between both surface density scalings persists and may even be exacerbated. Based on our comparison to observations in Figure~\ref{fig:rsig_obs}, we believe the relation between cluster radius and surface density is better described by our model. The discrepancy might be explained by the omission of galactic dynamics in the simulations of \citet{grudic_etal_2020}, whereas these play a central role in setting the mass scales of clouds and cluster-forming clumps in our model. The weak dependence on the cloud mass in equation~(\ref{eq:grudic}) may also help decrease the difference, because this would require $M_{\rm cloud}\propto\Sigma_{\rm cloud}^{5/2}$ and thus $R_{\rm cloud}\propto M_{\rm cloud}^{3/10}$, which is not too far off the $R_{\rm cloud}\propto M_{\rm cloud}^{1/2}$ typically found in observations \citep[e.g.][]{romanduval_etal_2010,rosolowsky_etal_2021}. Indeed, our model's dependence on the galactic environment implies $M_{\rm cloud}\propto\Sigma_{\rm cloud}^3$ for a fixed angular velocity of the rotation curve \citep[see Section~\ref{sec:methodology} and][]{reina-campos_kruijssen_2017}. Accounting for the approximate scaling between surface density and angular velocity in nearby galaxies \citep[e.g.][]{krumholz_mckee_2005} might weaken this to $M_{\rm cloud}\propto\Sigma_{\rm cloud}$. To definitively compare our predicted environmental dependence of the mass-radius relation to the numerical results of \citet{grudic_etal_2020}, the simulations would need to be rerun in the context of the galactic environment. Finally, our model omits an explicit dependence on the metallicity as in the simulations of \citet{grudic_etal_2020}. However, this dependence is so weak that we consider it a negligible effect within the uncertainties associated with our simple model.

\subsection{Model limitations}
\label{subsec:discussion_limits}
Our model implicitly neglects several complicating pieces of physics that may be relevant to setting the initial radii of star clusters. \cite{gieles_renaud_2016} showed that impulsive tidal shocks (by e.g.\ passing molecular clouds) can \textit{decrease} cluster radii, competing with gradual expansion driven by two-body relaxation. Therefore, in reality, the dynamically evolved curves shown in Figs. \ref{fig:param_space}-\ref{fig:cmz} may serve as upper limits to the expected cluster radius. Moreover, given the inherent stochasticity in the time for a cluster to encounter a GMC, shocks almost certainly contribute (downward) scatter in the observed MRR. This may explain why our predicted MRR appears to have a slightly higher normalization and lower scatter than observed.

Another complication that we have neglected in this work is the role of stellar feedback in restructuring the potential of the gas cloud, beyond our simple approximation of adiabatic gas expulsion. Recent numerical simulations of star formation reveal that the interplay between various feedback sources (e.g.\ photoionization, radiation pressure, winds) is inherently complex. Moreover, their impact in simulations depends sensitively on the exact numerical implementation \citep{dale_etal_2005,roskar_etal_2014, grudic_etal_2018_elephant, kim_ostriker_2018,  li_etal_2019}. Therefore, a detailed accounting for feedback is well beyond the scope of this work. On the other hand, star formation occurs over a wide, continuous density distribution and bound clusters are expected to emerge from the highest density tail of this probability distribution \citep{elmegreen_elmegreen_2001, bressert_etal_2010, parker_meyer_2012, kruijssen_2012_cfe}. The importance of feedback can be roughly expressed as the ratio of the cloud self-gravity $F_g \sim G\Mcl/\Rcl^2$ and the feedback force $F_{\rm fb} \sim \Mstar\dot{P}_{\star}/\bar{m}$, where $\dot{P}_{\star}$ is the momentum deposition rate from feedback; the ratio of the two forces $F_g/F_{\rm fb} \propto \Sigma_{\rm cl}$, and hence at high densities feedback is expected to become ineffective  \citep{fall_etal_2010, murray_etal_2010, raskutti_etal_2016, ginsburg_etal_2016, grudic_etal_2018_feedbackfails, li_etal_2019}. We take the reasonable agreement between our simple model and observations as an encouraging hint in support of this point.

\section{Conclusions}
\label{sec:conclusions}
In this work, we explored the initial mass-radius relation of young stellar clusters. We summarize the key results of our work below:
\begin{enumerate}
    \item We develop a simple model for the initial mass-radius relation (MRR) of stellar clusters. The model assumes stellar clusters form in self-gravitating clumps within GMCs. We relate the properties of the cluster to those of the cloud and in turn relate the cloud properties to those of the large-scale galactic environment. For a fixed galactic environment, the predicted initial MRR is one of constant surface density, $\Rstar \propto \Mstar^{1/2}$ [equation~(\ref{eqn:rm})].
    \item We also predict that the normalization of the MRR varies with the surface density of gas in the ambient ISM as $\Rstar \propto \sigg^{-1/2}$. We test this model prediction against observed cluster populations across nearly three orders of magnitude in gas surface density and find excellent agreement (Fig.~\ref{fig:rsig_obs}).
    \item We show that the \textit{global} MRR slope can be shallower than the local MRR slope (in our case, $\Rstar \propto \Mstar^{1/2}$). At the high-mass end of the MRR, this is because massive clusters can \textit{only} form in high gas-density environments, which in turn decreases the MRR normalization at the high-mass end. At the low-mass end dynamical evolution quickly shallows the slope and washes out memory of the initial MRR. Combining the two effects yields an MRR of  roughly $\Rstar \propto \Mstar^{1/3}$ (Fig.~\ref{fig:slope_test}), similar to the observed trend.
    \item At the low-mass end of the MRR, dynamical evolution washes out memory of the initial cluster properties. The high-mass end of the observed MRR is relatively unaffected by stellar dynamics and the initial MRR is therefore preserved. 
    \item We test the model against the young cluster populations of the disc galaxies M31, M51, M83, and M74 and find generally good agreement throughout (Fig.~\ref{fig:discs}), although the overall normalization predicted by our model in these galaxies appears to be too high. We also test the model in the nuclear starbursting regions of M83, NGC 253, and the Milky Way's Central Molecular Zone (Figs. \ref{fig:starbursts}-\ref{fig:cmz}). We find excellent agreement in these environments.
\end{enumerate}
The analytic nature of our model makes it highly suitable for inclusion in numerical simulations of galaxy formation that track the formation and evolution of the stellar cluster population. In addition, having access to a simple, physically-motivated model for the MRR enables testable predictions for future observations of proto-GCs at high redshift and the modelling of GCs as hosts of LIGO gravitational wave sources. In addition to these potential applications, our model represents a first step towards a understanding the MRR.

\section*{Acknowledgements}
We thank Bruce Elmegreen, Mark Gieles, Oleg Gnedin, Mike Grudic, Munan Gong, Mark Krumholz, Hui Li, Marta Reina-Campos, Joel Pfeffer, and Tom Zick for useful exchanges, and the anonymous referees for constructive reports which significantly improved the quality of this work. We thank M\'{e}lanie Chevance and the PHANGS collaboration for constructing and sharing gas density profiles of M51 and M74, Andreas Schruba for sharing gas density profiles of M31, and Jenna Ryon, Angela Adamo and Cliff Johnson for sharing their data on young cluster populations in M51 and M31. Finally, we thank Marta Reina-Campos for sharing her literature compilation of galactic gas density profiles and the code for her maximum cluster mass model. NC thanks the entire MUSTANG group for their hospitality during the course of this work. NC and JMDK gratefully acknowledge support from Sonderforschungsbereich SFB 881 ``The Milky Way System'' (subproject B2) of the German Research Foundation (DFG). JMDK gratefully acknowledges funding from the DFG in the form of an Emmy Noether Research Group (grant number KR4801/1-1) and the DFG Sachbeihilfe (grant number KR4801/2-1), as well as from the European Research Council (ERC) under the European Union's Horizon 2020 research and innovation programme via the ERC Starting Grant MUSTANG (grant agreement number 714907). This work made use of the \textsc{scipy} \citep{scipy} and \textsc{matplotlib} \citep{matplotlib} packages.

\section*{Data availability}

No new data products were generated for this article. The used data are available upon reasonable request to the corresponding author.




\bibliographystyle{mnras}
\bibliography{sizes/gc_nick} 



\appendix



\bsp	
\label{lastpage}
\end{document}